\documentclass[useAMS]{mn2e}
\usepackage{epsfig}

\def\arcsec{\hbox{$^{\prime\prime}$}}

\newcommand*{\ngc}{NGC\,253}
\newcommand*{\cxou}{CXOU\,J004732.0-251722.1}

\title[WR XRB in NGC~253]{A new candidate Wolf-Rayet X-ray binary in NGC~253}

\author[Maccarone et al.]{Thomas J. Maccarone$^{1}$\thanks{email:thomas.maccarone@ttu.edu},
Bret D. Lehmer$^{2,3}$, J.C. Leyder$^{3,6}$, Vallia Antoniou$^{4,7}$,\newauthor Ann Hornschemeier$^{3}$, Andrew Ptak$^{2}$, Daniel Wik$^{3}$, Andreas Zezas$^{5}$\\
$^{1}$ Department of Physics, Texas Tech University, Lubbock TX, 79409, USA \\
$^2$ The Johns Hopkins University, Homewood Campus, Baltimore MD, 21218, USA\\
$^3$ NASA Goddard Space Flight Center, Code 662, Greenbelt MD, 20771, USA\\
$^4$ Department of Physics and Astronomy, Iowa State University, 12 Physics Hall, Ames IA, 50011, USA\\
$^5$ Physics Depatment, University of Crete, Heraklion, Greece\\
$^6$ European Space Agency, European Space Astronomy Centre, PO Box 78, E-28691, Villanueva de la Ca\~nada, Madrid, Spain\\
$^7$ Harvard-Smithsonian Center for Astrophysics, 60 Garden Street, Cambridge MA 02138, USA\\}

\begin{document}
\def\ltsim{\mathrel{\rlap{\lower 3pt\hbox{$\sim$}}
        \raise 2.0pt\hbox{$<$}}}
\def\gtsim{\mathrel{\rlap{\lower 3pt\hbox{$\sim$}}
        \raise 2.0pt\hbox{$>$}}}

\date{}

\pagerange{\pageref{firstpage}--\pageref{lastpage}} \pubyear{}

\maketitle

\label{firstpage}

\begin{abstract}
We have discovered a persistent, but highly variable X-ray source in
the nearby starburst galaxy NGC~253.  The source varies at the level
of a factor of about 5 in count rate on timescales of a few hours.
Two long observations of the source with Chandra and XMM-Newton show
suggestive evidence for the source having a period of about 14-15
hours, but the time sampling in existing data is insufficient to allow
a firm determination that the source is periodic.  Given the
amplitude of variation and the location in a nuclear starburst, the
source is likely to be a Wolf-Rayet X-ray binary, with the tentative
period being the orbital period of the system.  In light of the fact
that we have demonstrated that careful examination of the variability
of moderately bright X-ray sources in nearby galaxies can turn up
candidate Wolf-Rayet X-ray binaries, we discuss the implications of
Wolf-Rayet X-ray binaries for predictions of the gravitational wave
source event rate, and, potentially, interpretations of the events.
\end{abstract}

\begin{keywords}
X-rays:binaries -- galaxies:individual:NGC~253 --  galaxies:starburst -- stars:Wolf-Rayet
\end{keywords}

\section{Introduction}
Wolf-Rayet X-ray binaries are a subclass of high mass X-ray binaries
where the donor star is a Wolf-Rayet star, rather than an ordinary
high mass star.  These systems present two golden opportunities to
study extreme physics using binary stars that are not seen in most
other classes of X-ray binaries.  Firstly, the archetypal Wolf-Rayet
X-ray binary, Cygnus X-3 (see e.g. van Kerkwijk et al. 1996), is also
the first clear example of an accretion-powered stellar mass system
that produces high energy gamma-rays (Tavani et al. 2009), with the
prevailing interpretation being that the system is a strong high
energy emitter due to interactions between its jet and its strong
stellar wind (e.g. Dubus et al. 2010).

Secondly, Wolf-Rayet X-ray binaries represent one of the best
candidate sources of gravitational wave source progenitors in the
Universe (e.g. Belczy\'nski et al. 2013).  As systems with orbital
periods of $\ltsim$ 1 day, they are more robust to remaining bound
after the second supernova than are very wide high mass X-ray
binaries.  They may thus lead to double black holes with merger
timescales shorter than a Hubble time, meaning that detection of these
systems can give real insight into the expected rate of mergers to be
seen with experiments like Advanced LIGO (Belczy\'nski et al. 2013).

To date, there are three convincing candidate Wolf-Rayet X-ray
binaries known -- Cygnus X-3 (van Kerkwijk et al. 1996), NGC~300~X-1
(Carpano et al. 2007), and IC~10~X-1 (Clark \& Crowther 2004; Bauer \&
Brandt 2004).  There is also a fourth object, SS~433, which has
sometimes been suggested to be a Wolf-Rayet X-ray binary, but which is
more likely to be mimicking such a system due to a strong disk wind
(Fuchs et al. 2006).  It is, however, worth remembering that these
systems are quite short-lived, so that for every one we see, many more
have likely been formed and died in the past.  Thus while these
objects may be rare, they may still represent an extremely important
progenitor channel for the formation of double black hole and black
hole/neutron star binary systems which may merge on a Hubble time.  In
this paper, we report the discovery of a new candidate Wolf-Rayet
X-ray binary in the galaxy NGC~253, and discuss its implications for
understanding the double compact object merger rate.  We note that in
this paper, we take the term Wolf-Rayet star to mean any massive star
which has completely lost its hydrogen envelope. We note that some
authors use a more restrictive definition, which distinguishes between
Wolf-Rayet stars and other classes of naked helium stars based on
details of optical and infrared spectroscopy -- see e.g. Linden et
al. (2012) and references within for a further discussion.

The structure of this paper is as follows: in section 2, we present a
discussion of the observations used and the data analysis methodology;
in section 3, we discuss the results of the X-ray data analysis; in
section 4, we discuss the interpretation of this object as a
Wolf-Rayet X-ray binary; in section 5, we outline a framework for
converting detected Wolf-Rayet binaries into compact object merger
rates; and in section 6, we summarize our conclusions.

\section{Observations and data analysis}
We examined all public observations of \ngc\ from Chandra and
  XMM-Newton for which the integrations are longer than 5000~s of
  exposure time, and we also used three proprietary Chandra
  observations from 2012. The data are summarized in
  Table~\ref{datatable}.  In this section, we discuss which data are
  used, and the procedures used for the analysis, while in the next
  section, we discuss the specific results.

This project began with the discovery of a strongly variable
  source, \cxou, in the proprietary Chandra observations taken in 2012
  (observations M--O in Table~\ref{datatable}). Following that discovery, we
  examined the data from the other available X-ray observations, in
  order to see if that variation was usual, and if a period could be
  identified.

\subsection{Chandra data analysis}
The Chandra observations from before 2004 were all obtained
  with ACIS-S, while those from 2012 were obtained with ACIS-I.  The
  data were analysed using the Chandra Interactive Analysis of
  Observations (CIAO) software, version 4.3. The extraction region
  used is a 2.5\arcsec\ circle centered on \cxou. We used a 2--7~keV
  bandpass for Chandra because the lower energies are heavily affected
  by diffuse emission from gas in NGC~253 itself.  The Chandra image,
  along with the XMM-Newton images, from the respective longest
  exposures, are show in Figure 1.

\subsection{XMM-Newton data analysis}
In order to check if similarly large variations from
  \cxou\ had been detected in previous \textit{XMM-Newton}
  observations of NGC\,253, the longest individual exposure was
  selected: \texttt{ObsID 0152020101} was taken in June 2003, and
  lasted about 141~ks. Those data were taken in Full Frame mode, with
  the medium filter for the MOS, and the thin filter for the pn. The
  data were reduced using the XMM-Newton Science Analysis Software
  (SAS) version 12.0.1, with standard parameters and methods. After
  data reduction, exclusion of the intervals affected by high particle
  background thanks to the application of the appropriate good time
  intervals (GTIs), and selection of the energy range of interest
  (0.2--10.0~keV), the exposure time was reduced to about 110~ks.

The XMM-Newton lightcurves for \cxou\ were extracted from the three
detectors (MOS1, MOS2, and pn) using a circular aperture with a
diameter of 5\arcsec\ and centered on the best position derived from
the \textit{Chandra} data. We note that this region is much smaller
than the point spread function (PSF) of XMM-Newton (encompassing
roughly 30\% of the photons emitted by the source for the pn, and
about 40\% for the MOS). The 5\arcsec\ diameter was chosen to maximize
the fraction of PSF included while minimizing contamination from the
numerous nearby sources; any larger radius would start to include
unrelated emission from the nucleus. All XMM-Newton lightcurves
presented in this paper were corrected for background, using an
extraction region located far from the nucleus. Finally, the source
lightcurves are corrected for various effects affecting the detection
efficiency and for effects affecting the stability of the detection
within the exposure using the \texttt{epiclccorr} tool; this
correction accounts for the PSF effects.

We note that the extraction region chosen is relatively small compared
to the pixel size of the XMM-Newton instruments (1.1\arcsec\ for the
MOS, 4.1\arcsec\ for the pn). However, the events generated by the SAS
are randomly re-distributed over the instrument pixel into sub-pixels
of 0.05\arcsec. Spatial coordinates are always randomized to avoid
Moiree patterns. Thus, extracting the photons from a small region
leads to a correct data products (\textit{i.e.} spectra and
lightcurves) thanks to the randomization process and the use of
sub-pixels, regardless of the position of that extraction region
inside the natural instrument pixel. Of course, all images produced
from those event lists are rebinned to match the physical detector
pixel sizes.

The Chandra and XMM-Newton pn light curves are presented in figure
\ref{lightcurve_figure}.  The MOS data have lower signal to noise
because of their smaller collecting areas than the pn, and poorer
background subtraction than Chandra, but show results consistent with
those from the pn data.  The MOS data are shown in figure
\ref{mos_lc_figure}.

\begin{table*}
\begin{tabular}{llllll}
\hline
Observation&Date&Start time& Exp. time (sec) & Satellite& ObsID\\
\hline
A&16 December 1999& 12:11:02& 13990& Chandra& 969\\ 
B&27 December 1999& 2:19:05& 43610& Chandra& 790\\ 
C&3 June 2000& 5:15:36& 60809 & XMM-Newton& 0125960101\\
D& 4 June 2000& 5:07:01& 17509 & XMM-Newton& 0125960201\\
E&14 December 2000 & 09:00:02& 33732& XMM-Newton& 011900101\\
F&20 June 2003&08:13:24 &140799 &XMM-Newton& 0152020101\\
G&19 September 2003& 12:50:20& 82550& Chandra& 3931\\
H&16 December 2005& 20:14:55& 23211& XMM-Newton& 0304851101\\
I&2 January 2006& 7:46:27& 11811& XMM-Newton& 0304850901\\
J&6 January 2006& 4:12:06& 11846& XMM-Newton& 0304851001\\
K&9 January 2006& 18:46:09& 19918& XMM-Newton& 0304851201\\
L&11 January 2006& 1:59:10& 20916& XMM-Newton& 0304851301\\
M&2 September 2012& 5:59:52& 19720&Chandra& 13830\\
N&18 September 2012& 3:37:53& 19720&Chandra& 13831\\
O&16 November 2012& 1:04:25& 19720&Chandra& 13832\\
\end{tabular}
\caption{The observations used for this project -- the columns are the start date for the observation; the start time for the observation, the exposure time, the satellite used, and the Observation ID number.}
\label{datatable}
\end{table*}

%

\section{Results}

Some early results for our Chandra-NuSTAR observations of NGC~253,
focused on its variable nucleus, have already been reported in Lehmer
et al. (2013).  Our analysis of the Chandra data taken during 2012 for
that project show CXO~J004732.0-25172.1 with about 100 source counts
on September 2, 60 source counts on 18 September, and an upper limit
of 8 source counts on 16 November, with all counts reported in the 2-7
keV band.  The location of the source, in J2000 coordinates, at right
ascension of 0:47:32.0 and declination of -25:17:22.1 is 17 arcseconds
from the center of NGC~253 (Veron-Cetty \& Veron 2010).

\begin{figure*}
\includegraphics[width=7 in]{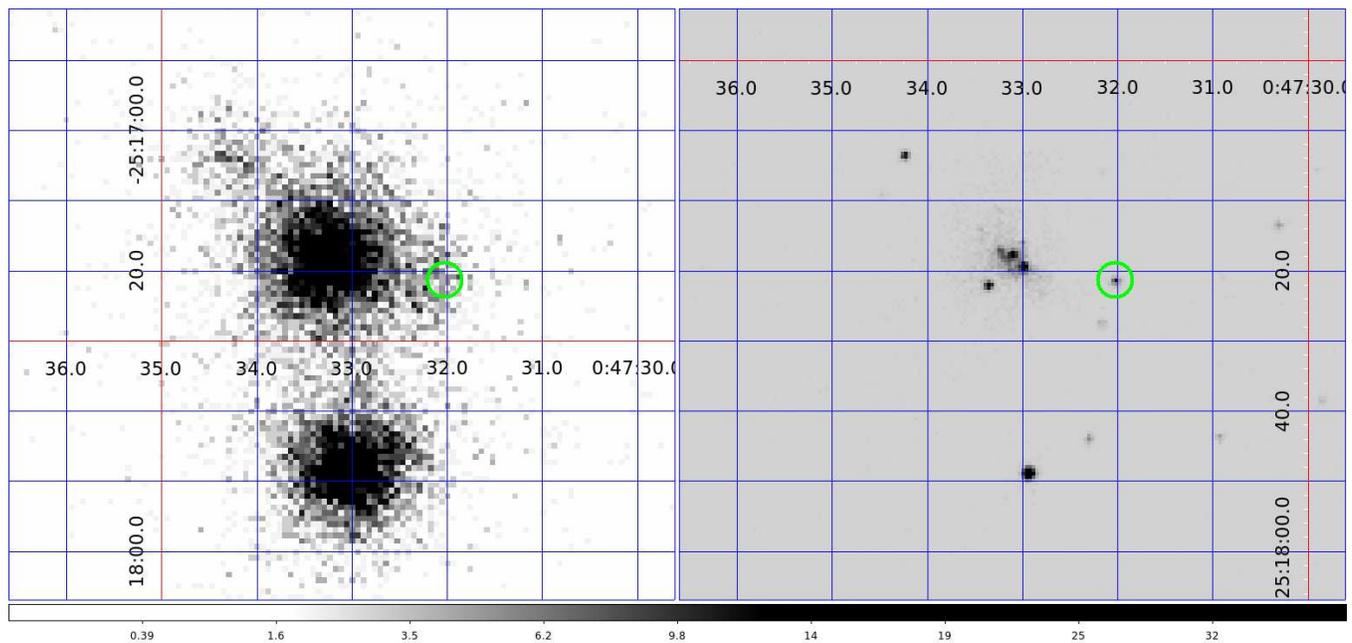}
\caption{Left: the XMM image of the same region, with the WCS matched
  between the two images, taken during observation 0152020101
  (Observation F).  Right:The Chandra image from observation 3931 of
  the candidate Wolf-Rayet X-ray binary (Observation G).  The green
  circle has a 2.5'' radius around the brightest pixel of the
  Wolf-Rayet binary candidate.  Both images show photons from 2-7 keV.
  Of particular importance to note is that there is no strong extended
  structure in the Chandra image near the position of the bright
  source, so that any emission there must be from point sources.  The
  images have been cleaned to remove flaring background periods.}

\end{figure*}

\begin{figure*}
\includegraphics[width=7 in]{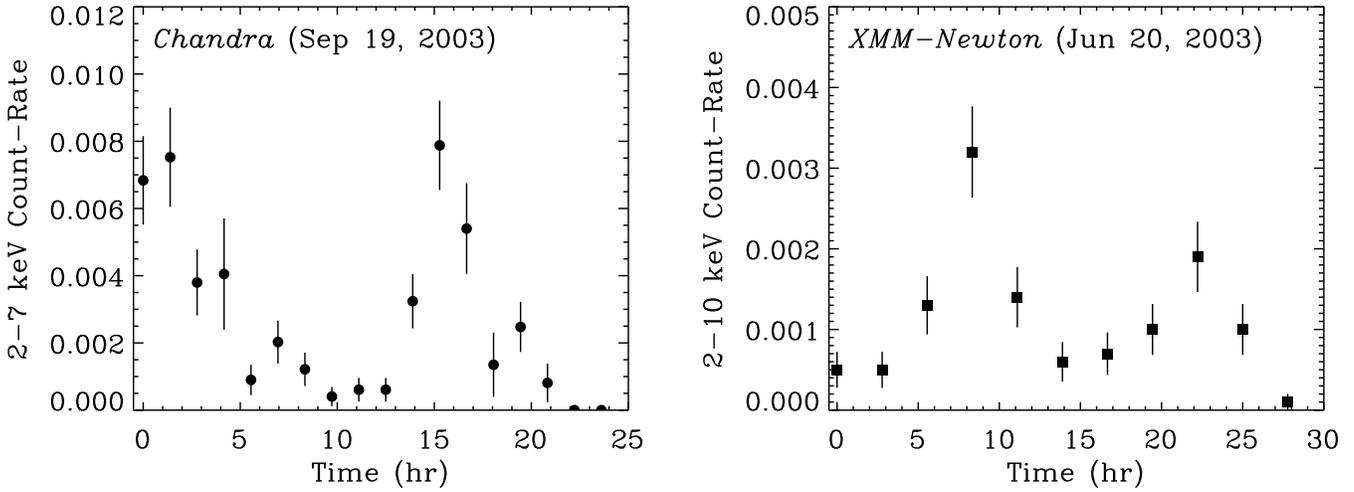}
\caption{Left: The Chandra light curve for the source.  Right: the
  XMM-Newton pn light curve for the source.}
\label{lightcurve_figure}
\end{figure*}

\begin{figure*}
\includegraphics[width=2.5 in,angle=-90]{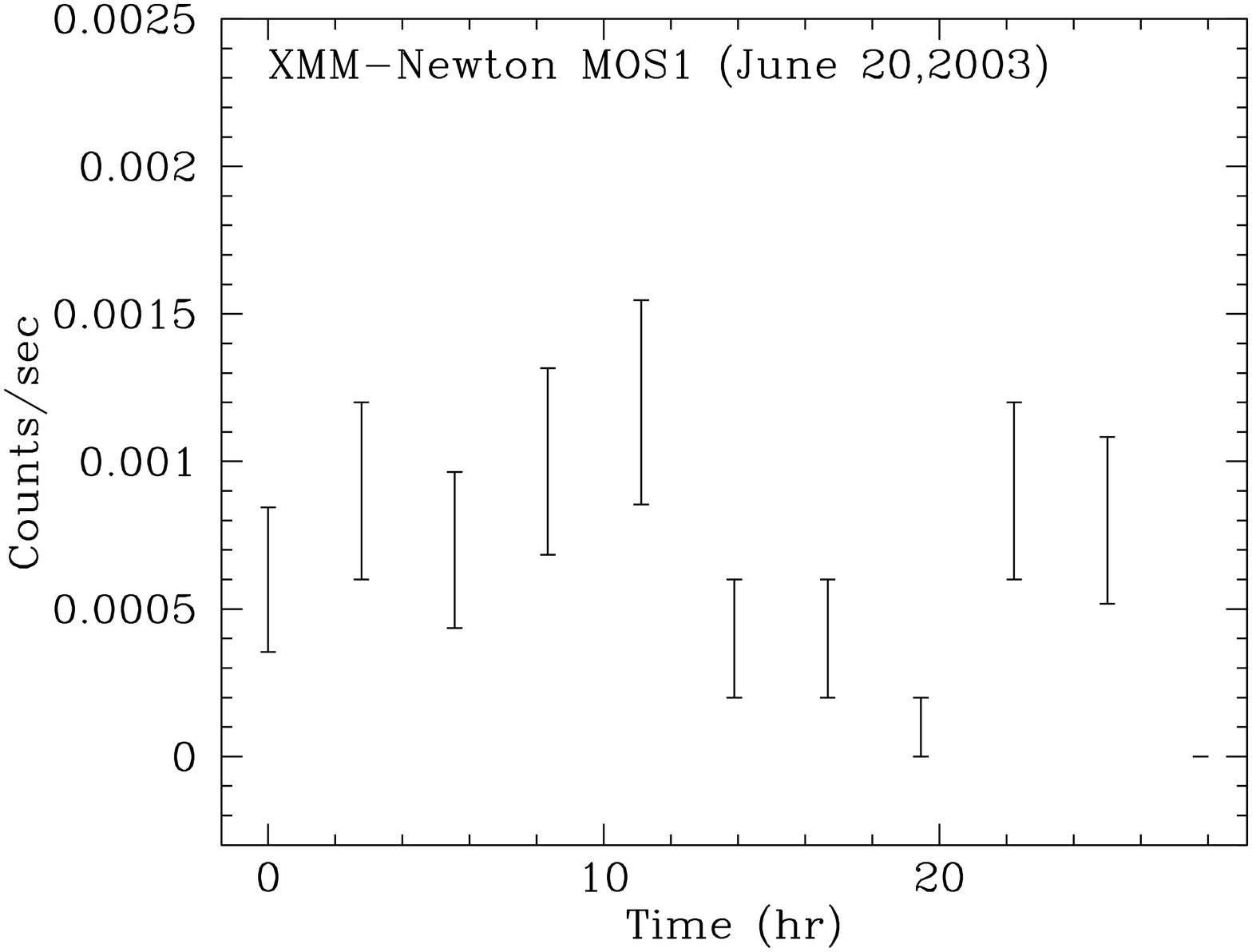}
\includegraphics[width=2.5 in,angle=-90]{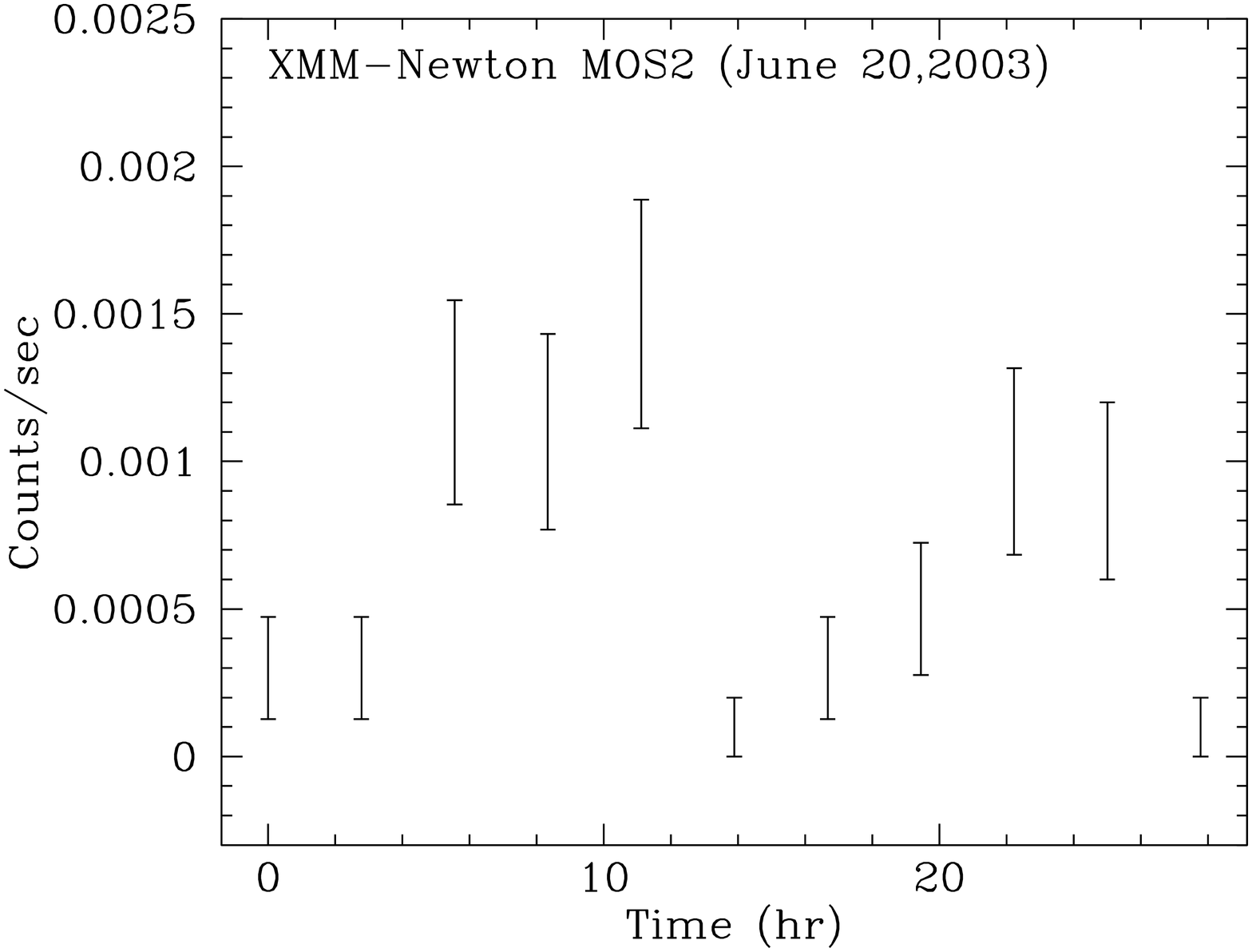}
\caption{The EPIC MOS light curves for observation 0152020101
  (i.e. the same observation for which the pn light curve is plotted
  in figure \ref{lightcurve_figure}).  One can clearly see that the
  peaks and troughs in the light curve are in the same places as for
  the PN light curve.}
\label{mos_lc_figure}
\end{figure*}

Given this unusually fast variability, we examined the older
observations of the source to determine whether it was a transient.
We found it to be bright in several past observations.  We then
examined its intra-observation variability in the two longest
observations -- F and G in Table 1.  The source count rate varies by a
factor of about 8 on timescales of a bit more than half a day -- see
figure \ref{lightcurve_figure}.  The nature of the variability is
suggestive of a periodic system with a large amplitude of variability,
but given that fewer than two cycles are seen if the source is
periodic, no definitive statement can be made on the basis of the
Chandra data alone.  

We next extracted light curves with 5000 second binning from all the
observations and computed a Lomb-Scargle (Lomb 1976;Scargle 1982)
periodogram of the data.  The strongest candidate periodicity is at
7.57 hours, which may be either the first harmonic of the period, or
an alias of the first harmonic.  It might be expected that the
harmonics of the period will be stronger than the fundamental in a
periodogram because the light curve does not appear to be sinusoidal.
No strong signal, however, is found at 15.04 hours, arguing that if
the signal at 7.57 hours is the real first harmonic, that an alias of
the fundamental is much stronger than the fundamental itself.  The
poor sampling of the data makes it difficult to establish any of the
weaker peaks in the data as a genuine periodicity.  More data, with a
sampling pattern more appropriate to the expected 14-15 hour period,
are needed in order to establish clearly whether the source is
strictly periodic, and what the period is.  We also note that we
searched the NuSTAR data at the source position for possible
periodicities as well, but found only periods and aliases of periods
related to the satellite orbit.

A natural explanation for a recurrent transient in a young stellar
population (which the central region of NGC~253, being a starburst,
is) would be for it to be a Be X-ray binary.  However, the candidate
period of about 15 hours is far too long to be a neutron star pulse
period, and far too short to be the orbital period for a Be X-ray
binary (i.e. a system with a neutron star or black hole accreting from
the equatorial wind of a Be star).  The orbital periods of Be X-ray
binaries tend to be a few months, and their frequent, type I,
outbursts tend to last about 0.2-0.3 orbital periods, and to peak at
no more than about $10^{37}$ erg/sec, while their less frequent Type
II outbursts tend to show smooth light curves on timescales of days
(see e.g. Reig \& Nespoli 2013).  In detail, then, the system's light
curve does not resemble that of a Be X-ray binary, and some
alternative form of highly variable source that can exist in a young
stellar population is needed.  The candidate period and the large
amplitude of variablity are in line with the empirical results from
known Wolf-Rayet X-ray binaries (e.g. Carpano et al. 2007; Bauer \&
Brandt 2004).

\subsection{The X-ray spectrum}

Given that the source has relatively few counts, detailed X-ray
spectroscopy cannot be done.  The 19 September 2003 observation has
the largest total number of counts within 2.5'' -- the XMM-Newton
observations would normally provide better spectroscopy than the
Chandra observations, but for this case, only the core of the PSF can
be used for the XMM-Newton data analysis, due to the crowding near the
source location.

We extract the Chandra spectrum of the source using the CIAO
specextract tool and a 2.5'' source region, and group the channels
into bins of at least 15 counts per channel.  We also use an
off-source background region, centered at RA of 00:47:37.1 and Dec of
-25:17:08.9, with a radius of 29''.  The background region was chosen
to be at a similar surface brightness to the region around our source
of interest while also not including any bright sources. Because there
are too few counts in the source-free parts of the local background
region to make a local background estimate, there will remain the
possibility that the spectral fits are affected by the detailed shape
of the background spectrum. We first fit the spectra from 0.5-7.0 keV,
accounting for the full range over which there are many counts and the
Chandra response matrix is well-calibrated, and then repeat the fits
from 2.0-7.0 keV to minimize the effects from contamination from the
diffuse background -- given the moderate number of total counts, we
believe this is the most robust way to handle the possibility of an
unusual local background.

We consider several spectral models.  First we consider two of the
simplest models commonly used to fit the spectra of accreting sources
-- a power law and a disk blackbody (Mistuda et al. 1984).  In both
cases, the foreground absorption is fixed to the Galactic column
density of $1.4\times10^{20}$ cm$^{-2}$ (Dickey \& Lockman 1990). We
use the phabs component in XSPEC to describe the absorption.  In both
cases, the fit is formally unacceptable, and produces physically
unlikely parameter values -- for the power law, the photon index
$\Gamma$ is 0.45 and $\chi^2/\nu=45.2/23$.  For the disk blackbody,
the inner disk temperature is 17.2 keV and $\chi^2/\nu=51.0/23$.
Uncertainties cannot be computed within XSPEC for sources with
$\chi^2/\nu>2$, so these values are presented without error bars. The
source has 394 counts from 0.5-7.0 keV and 265 counts in the 2-7 keV
energy range.

This is perhaps not surprising, since there is likely to be additional
absorption both by the wind of the donor star and by the interstellar
medium of NGC~253.  We then re-fit the spectrum with the same two
models, allowing $N_H$ to float freely.  For the disk blackbody we
find $N_H=8.0^{+0.37}_{-0.30}\times10^{21}$cm$^{-2}$,
$k_BT=2.2^{+1.4}_{-0.6}$ keV, with $\chi^2/\nu=25.5/22$, and hence a
null hypothesis probability of 0.27.  For the power law, we find
$N_H=1.0^{+0.5}_{-0.4}\times10^{22}$cm$^{-2}$ and
$\Gamma=1.37^{+0.45}_{-0.41}$, with $\chi^2/\nu=28.2/22$, so there is
a null hypothesis probability of 0.17.  The 90\% confidence interval
is given for all parameters.  

\begin{figure}
\includegraphics[width = 2.7 in, angle=-90]{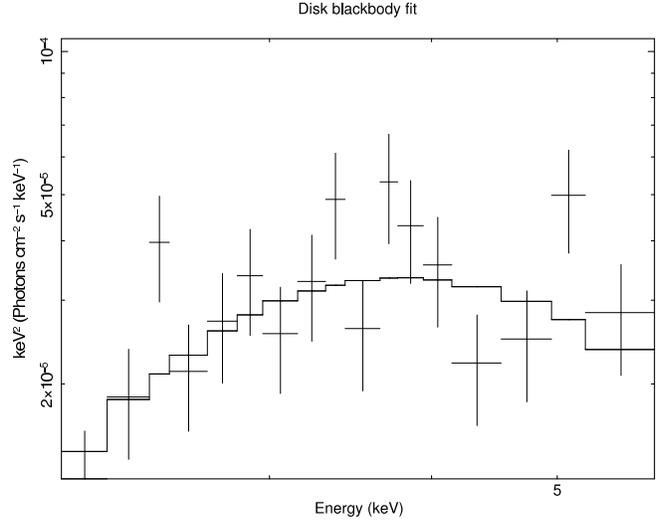}
\caption{The unfolded spectrum of our candidate Wolf Rayet X-ray binary, fitted with a disk blackbody model.  One can see that the model describes the data well, and that more sophisticated models are unlikely to be well constrained given the quality of the data.}
\label{diskbb}
\end{figure}

When fitting from 2-7 keV, the power law best-fit values are
$N_H=5.6^{+3.3}_{-2.7}\times10^{22}$cm$^{-2}$ and
$\Gamma=3.4^{+1.4}_{-1.2}$, with $\chi^2/\nu=17.0/14$, so there is a
null hypothesis probability of 0.26, and the disk blackbody fits yield
$N_H=3.1^{+2.2}_{-1.9}\times10^{22}$cm$^{-2}$,
$k_BT=1.2^{+0.7}_{-0.3}$ keV, with $\chi^2/\nu=18.1/14$, and hence a
null hypothesis probability of 0.20.  The absorption columns are
marginally larger than those inferred for the fits including the
0.5-2.0 keV band.  The disk blackbody model, which we regard as the
most likely of the bunch for reasons discussed below, is plotted in
figure \ref{diskbb}.

While these models are not likely to provide full descriptions of the
data, given the variability, regardless of the interpretation, more
detailed modeling is not justified by the data.  One broad conclusion
is well-justified by the data -- that significant foreground
absorption in excess of the Galactic foreground is required.  Beyond
this, the results for the fits made from 2-7 keV argue mildly for the
source to be a black hole in a canonical soft state (i.e. a disk
blackbody with $k_BT\sim$1 keV), rather than a black hole in a hard
state (a power law with $\Gamma\sim$$1.5-1.7$) or a high magnetic
field accreting X-ray pulsar (a power law with $\Gamma\sim$$0.5-1.0$),
because the paramter values for the soft state black hole case fit
well to the data, while the fitted parameter values do not match those
expected for the other two cases.  Also, the model for the disk
blackbody model corresponds to an unabsorbed flux of
$1.4\times10^{-13}$ ergs/sec/cm$^2$ from 0.5-8 keV, which gives a
luminosity of $1\times10^{38}$erg/sec.  Such a luminosity is a
reasonable luminosity to expect for a soft state black hole, but it
would be unusual for a hard state black hole, assuming that the black
hole mass is about $10M_\odot$.  The typical state transition
luminosities between the hard state and soft states for black holes
typically occur at about 2\% of the Eddington luminosity (Maccarone
2003), with only short-lived hard states which are brighter and occur
during outbursts as a hysteresis effect (Maccarone \& Coppi 2003).

We also consider the possibility that the source has a neutron star
accretor with a low magnetic field, and hence behaves like the
Galactic ``Z-sources'' -- the low magnetic field neutron stars
accreting near the Eddington rate.  We take a single example of a
spectrum of the Z-source LMC X-2 from Lavagetto et al. 2008) and
determine whether the model is consistent with the data for this
source.  We use the parameter values for the disk blackbody +
blackbody model from Lavagetto et al. (2008), and freeze our blackbody
temperature to the 1.54 keV value they find, and freeze our disk
blackbody temperature to the 0.815 keV value they find.  Since they
use XMM-Newton data with a broader bandpass than our Chandra data
cover, there is no risk that the extrapolation of their spectrum to
our energy range will result in an unacceptable fit because of a poor
model.  We find that, when varying only the normalizations of the
model component and the foreground $N_H$, an acceptable fit is
obtained, with $\chi^2/\nu$=17.2/14, and a null hypothesis probability
of 0.25.  While a low magnetic field for a neutron star in such a
young population is unlikely, the Galactic source Circinus X-1, for
example, seems to be such an object (Boutloukos et al. 2006), so the
possibility remains viable (e.g. the LMC, which is the galaxy for
which we took our prototype source LMC X-2, is also a star forming
galaxy).

\section{Interpretation as a Wolf-Rayet X-ray binary}

While a periodicity cannot be clearly established given the lack of
intensive sampling of the source's variability, we still argue that
the most likely explanation for the observed variability is that it is
the orbital modulation of the X-ray flux due to the changes in the
absorption from the stellar wind of a Wolf-Rayet X-ray binary.

The location of the source in a star-forming galaxy suggests that the
source is likely to be some kind of high mass X-ray binary,
particularly since the presence of a supernova remnant within 2''
(projected distance of 25 pc) of the source (Collison et al. 1994)
indicates that it is in a region of current active star formation.  We
thus focus on high mass X-ray binary interpretations for the source,
but also consider low mass X-ray binary possibilities later.  We
cannot make any definitive statements about the nature of the source's
optical counterpart because there are not any optical or infrared data
which are deep enough and have astrometry sufficiently well-tied to
the X-ray data.

If the variations are periodic, then they almost certainly represent
the orbital period of a Wolf-Rayet X-ray binary.  Large amplitude
periodicities are generally not found in accreting black hole systems,
except on the orbital timescale.  Quasi-periodicities have been
suggested in several systems to take place on timescales of a few tens
of orbital periods (e.g. Smale \& Lochner 1992), with the most common
explanation being modulation due to precession of a warped accretion
disk (e.g. Maloney \& Begelman 1997; Wijers \& Pringle 1999).  Some
neutron star systems show strong periodic modulations as X-ray
pulsars; a pulse period of 14-15 hours coupled with a peak luminosity
of $10^{38}$ ergs/sec would require a magnetic field several orders of
magnitude larger than any seen to date (e.g. Ghosh \& Lamb 1978), and
can be taken as unlikely.

If the periodicity is real and orbital, and the system is a high mass X-ray
binary, then the donor star must be a Wolf-Rayet star.  The density of
the donor star of a Roche-lobe overflowing system is solely a function
of its orbital period:

\begin{equation}
\rho = 107 P_{orb,h}^{-2} {\rm g/cm}^3,
\end{equation}
from combining the Paczy\'nski (1971) formula for the Roche lobe size
with Kepler's third law (see e.g. Knigge et al. 2011).

The mass radius relation of main-sequence stars above 1.66 $M_\odot$
is given by:
\begin{equation}
{\rm log} R=0.12 + 0.55 {\log} M,
\end{equation}
with $R$ and $M$ in solar units (Demircan \& Kahraman 1991).  We can then see that the
density of a $8 M_\odot$ star will be about 0.1 g/cm$^3$, while the
Roche lobe density of a star in a 14.5 hour orbit will be 0.5
g/cm$^3$, so the 8 $M_\odot$ star will overfill its Roche lobe rather
dramatically; for the period to be orbital, the system must either
have a relatively dense, and hence low mass donor star, or it must
have a donor star which has lost its envelope (i.e. a Wolf-Rayet
star).  While some low mass X-ray binaries do show modulation on their
orbital periods, these tend to be quite low amplitude unless the
systems are eclipsing, and the eclipsers tend to show sharp ingresses
and egresses due to the small X-ray emission regions (e.g. Wolff et
al. 2009).

Next, we can consider possibilities for aperiodic variability from the
system.  In this case, the key parameters remain the relatively long
timescale of variations and the large amplitude of variations.  We can
consider the variations in persistent systems with similar peak X-ray
luminosities.  These are largely the Z-sources.  Looking at the RXTE
All-Sky Monitor light curves for the two best-studied Z-sources, Sco
X-1, and Cyg X-2, we see that neither varies at all by more than a
factor of about 3.  Consulting the High Energy Astrophysics Virtually
Enlightened Sky (HEAVENS) database of INTEGRAL data (Walter et
al. 2010), we see also that the accreting pulsar Vela X-1 similarly
rarely varies by more than a factor of a few when out of eclipse --
and is, in any event, at a much lower luminosity than this object --
in fact, all Galactic accreting pulsars are (see Lutovinov et al. 2013
for a compilation).  It is also difficult to envision from empirical
data how a black hole system would show such strong variations on
these timescales -- while the accreting black hole system GRS 1915+105
shows similarly strong variability on timescales of hours
(e.g. Belloni et al. 2000; Rodriguez et al. 2008), it does so at
luminosities near the Eddington limit, and the variations may be due,
in part to e.g. thermal instabilities for systems highly affected by
radiation pressure (e.g. Szuszkiewicz \& Miller 1998).

Evidence for quasi-periodic oscillations in the Galactic high magnetic
field neutron star X-ray binary EXO 2030+375 has also been seen
(Klochkov et al. 2011).  The authors argue that the mechanism is the
formation of a ``reservoir'' of gas when the magnetosphere of the
neutron star truncates the accretion disk onto the neutron star
outside of, but close to, the neutron star co-rotation radius
(D'Angelo \& Spruit 2010).  The reservoir is then drained on a viscous
timescale once enough material has piled up to overwhelm the neutron
star's magnetic field.  This mechanism is unlikely to explain a
$\sim15$ hour periodicity for this source -- the quasi-periodicity
seen in ESO 2030+375 is about 7 hours, and scales linearly with the
spin period of the neutron star.  Given that this source appears to be
persistently very bright it is likely to be a Roche Lobe overflowing
HMXB, if it is a neutron star HMXB accretor at all, and then
correlations between orbital period and spin period for Roche lobe
overflowers would yield an expected spin period several times shorter
than the spin period of 40 seconds seen in EXO 2030+375 (Corbet 1986).

When comparing the behavior of CXOU~J004732.0-251722 with the
phenomenological behavior of Galactic X-ray sources, it appears most
likely that the variability seen is periodic and orbital; the chief
caveat would be the possibility that the source is behaving in an
unusual manner relative to the observed phenomenology of Galactic
sources, despite being in a luminosity range which is fairly common
for Galactic sources.  

We can also argue that the source is more likely to have a black hole
accretor than a neutron star accretor on the basis of its X-ray
luminosity -- Linden et al. (2012) performed a population study of
neutron star X-ray binaries with naked helium star donors and found
that for orbital periods of $\sim 1$ day the X-ray luminosities were
typically less than $10^{37}$ erg/sec.  While it may be possible
either that the mass loss rates are higher than predicted by the
Hurley et al. (2002) formalism used by Linden et al. (2012), or that
the accretion is more efficient than Bondi-Hoyle (1944), the
luminosity being so high is at least a point in favor of the idea that
the accretor is a black hole, and that the donor is a relatively
massive Wolf-Rayet star.  We thus conclude that, while we cannot
firmly establish either a periodicity in this source or the mass of
the compact object, the most likely explanation for all its properties
is that it is a black hole X-ray binary with a Wolf-Rayet companion.
Additional data sets which could help verify the nature of the source
would be a better sampled X-ray light curve, which could establish or
refute the periodicity, and a deep infrared spectroscopy campaign,
which could find moving emission lines from the donor star.

\section{Converting the observed Wolf-Rayet binaries into a merger rate}

Theoretical uncertainties of a factor of about 1000 exist in
predictions of the rate of which ground-based gravitational wave
detectors will discover sources (Abadie et al. 2010), motivating a
more empirical basis for making predictions.  Procedures have been
developed in the past for converting small numbers of observed systems
into empirical event rates in the case of using the observed double
neutron star systems to estimate event rates for double neutron star
mergers (see Kim, Kalogera \& Lorimer 2003 for a recent example; Clark
et al. 1979 for the first example).  For understanding neutron star
mergers on the basis of double neutron stars, one of the chief
uncertainties comes from understanding the selection effects in pulsar
surveys (see e.g. Narayan 1987).

For understanding the rate of double compact object mergers on the
basis of populations of high mass X-ray binaries, considerably greater
uncertainties are present.  In general, because the second supernova
has not yet taken place, one needs to make assumptions about the
velocity kick distributions of the second supernova explosions.
Additionally, the mass and radius of the donor star are often
difficult to measure, especially in the cases of Wolf Rayet stars,
where it is difficult to see down to the stellar photosphere through
the stellar wind.

For understanding the mergers of two neutron stars, the double neutron
stars themselves are obviously a superior sample of objects with which
to work.  For understanding mergers involving black holes, on the
other hand, the double neutron stars are obviously of no use, so one
is stuck with using either purely theoretical calculations
(e.g. Belczynski et al. 2007) or one must attempt to forge ahead with
observations of X-ray binaries, aware of the extra complications in
interpreting their populations, and the greater uncertainties
involved.  Eventually, of course, some pulsar/black hole binaries may
start to be discovered, which could allow for the same kind of
empirical approach used for double neutron stars to be applied to
neutron star-black hole binaries.  A hybrid approach can be used, as
well, which identifies families of models that are consistent with the
existing observations of X-ray binaries; some success has been
obtained in recent years in modelling the X-ray luminosities and
luminosity functions of late-type galaxies, suggesting that this
approach has great potential (e.g. Tzanavaris et al. 2013; Tremmel et
al. 2013).

A procedure can be outlined for how to deal with this conversion for
high mass X-ray binaries as follows:

\begin{enumerate}
\item Identify the binary, and estimate its parameters.
\item Follow the evolution of the binary using a binary evolution code
  until it becomes a double black hole or a black hole/neutron star
  binary.  Estimate its lifetime as an X-ray binary, $t_X$.
\item Compute the merger timescale, $t_{merge}$.  If the merger timescale is larger than a Hubble time, then the object can be ignored.  Otherwise continue.
\item Compute the radius out to which the object could be detected as a gravitational wave source, $r_{detect}$.
\item Compute the star formation rate per comoving volume,
  $\dot{\rho}_{tform}$, of the Universe at lookback time $t_{merge}$,
  using one of the compilations of cosmic star formation history
  (e.g. Lilly et al. 1996; Madau et al. 1996; Hopkins 2004), and
  divide by the local star formation rate (e.g. from the
  11~HUGS\footnote{This acronym stands for the 11 Mpc H$\alpha$ UV
    Galaxy Survey} compilation of Lee et al. 2009 which includes all
  star-forming galaxies out to 11 Mpc distance), to get the
  amplification factor $A$ due to the fact that the gravitational wave
  events seen now are produced at times when the cosmic star formation
  rate was quite different from that now.
\item Integrate out the total star formation rate, $\dot{\rho}_{detect}$ within $r_{detect}$.  For double black  hole mergers with LIGO, $r_{detect}$ will generally be large enough that one need not worry about the effects of being located within regions of large scale structure that are not representative of the mean properties of the Universe, and small enough that one need  not worry about the lookback time to the outer part of the volume; with Advanced LIGO, a somewhat more sophisticated calculation will be necessary to deal with the lookback time issue.
\item Calculate the total star formation rate of galaxies which have been searched effectively for high mass binaries with similar X-ray luminosity and orbital period to the source under consideration, $\dot{\rho}_{search}$.  Note that at the present time, the strong candidates all have $L_X=10^{38}$ ergs/sec and all have orbital periods less than about 1.5 days, so they probably all could have been detected in all the galaxies searched.
\end{enumerate}

Each detected high mass binary will then contribute an expectation
value of
$\left(\frac{A}{t_X}\right)\left(\frac{\dot{\rho}_{detect}}{\dot{\rho}_{search}}\right)$
per year of gravitational wave events.  A lower limit on the
gravitational wave source detection rate can then be estimated by
summing over the detected objects.  This analysis ignores some
possible additional factors -- e.g. the metallicity at which star
formation typically takes place at moderate to high redshift may
differ from the metallicity at which star formation takes place at the
present time.  Lower metallicity stars may leave behind higher mass
black holes (e.g. Belczynski et al. 2010).  Additionally, such an
analysis ignores binaries formed through other mechanisms, such as
those formed dynamically in globular clusters (e.g. Portegies Zwart \&
McMillan 2000) -- a possibility that seems even more likely now than
it did when first proposed, given the discoveries of black hole X-ray
binaries in both extragalactic (Maccarone et al. 2007) and Galactic
(Strader et al. 2012) globular clusters.

For simplicity, in this paper, we make an illustrative calculation in
which we will assume that the donor high mass stars collapse promptly
into black holes of $10 M_\odot$ at the current orbital period, and
that the black holes already present in these systems are of about $10
M_\odot$.  This is below the estimates made for the black holes in
IC~10~X-1 (Prestwich et al. 2007; Silverman \& Filippenko 2008) and
NGC 300~X-1 (Crowther et al. 2010), an issue we discuss in section
\ref{vanKerkwijk}.  This assumption is obviously extremely crude, as
it neglects the effects of orbital period evolution, possible second
common envelope phases of stellar evolution, and supernova kicks.  A
proper treatment of the problem would require running binary evolution
calculations starting from the present conditions -- which themselves
are presently poorly understood.  We emphasize that estimates done
without undertaking proper binary evolution calculations should not be
taken too seriously, and that the purpose of the calculation here is
simply to illustrate that single object detections can have
significant implications for theoretical estimates of gravitational
wave deteciton rates.

Let us consider, then, a system that begins in an orbital period of 14
hours, and has two black holes, each of $10 M_\odot$.  The semi-major
axis of the binary is then $5.6\times10^{11}$cm.  Pfahl et al. (2005)
provides a useful expression for $t_{merge}$, based on the work of
Peters (1964).  for circular binaries, this is:
\begin{equation}
t_{merge}=2.1 {\rm Myr} \left(\frac{P_b}{\rm hr}\right)^{\frac{8}{3}} \left(\frac{m_1+m_2}{10 M_\odot}\right)^{\frac{-2}{3}} \left(\frac{\mu}{M_\odot}\right)^{-1}
\end{equation}
where $m_1$ and $m_2$ are the masses of the binary components, and
$\mu$ is their reduced mass, $P_b$ is the binary's orbital period, and
$e$ is its eccentricity.  For eccentric binaries, one must multiply
through by an additional factor of $(1-e^2)^{\frac{7}{2}}$.  Then, if
we consider a circular initial binary, we get $t_{merge}$ of about 300
Myrs.  Given this lookback time, the local star formation rate for the
Universe can be used, so $A=1$.  The radius out to which LIGO should
be able to detect mergers of two 10 $M_\odot$ black holes is about
$90$ Mpc, so we can take about $8^3$ times the star formation rate of
the 11 HUGS survey as the estimate of the total enclosed star
formation rate -- so we can presume that LIGO is sensitive to about
$6000 M_\odot$/yr worth of star formation.  At the present time, there
are no indications that careful searches which would exclude a
Wolf-Rayet nature have been done for any galaxies other than those for
which detections have been reported.  We can thus take the 11 HUGS
star formation rates for the moderately distant galaxies for which
detections have been made, NGC~253, NGC~300, IC~10, and the Milky Way,
obtaining roughly $4M_\odot$/yr, with considerable uncertainty in the
star formation rate of the Milky Way.  We find, then that based on our
new NGC~253 system alone, the birth rate of gravitational wave sources
is only about $2\times10^{-3}$ per year, assuming $t_X\approx{10^6}$
years.  However, enclosing a volume a factor of about 4000 larger, as
Advanced LIGO should do, would bring the event rate up to $\sim10$ per
year, based on this object alone.  Furthermore, we note that systems
with orbital periods a factor of 2.5 longer would have $t_{merge}$
about 11 times as long, meaning that they would have to be formed
about 3-4 Gyrs before the black holes merge, bringing them back to
$z\sim0.3$ if they merger locally, or $z\sim1$ for sources detected
near the edge of Advanced LIGO's distance range.  This would then
allow an additional enhancement in the event rate due to the
significantly higher star formation rate at $z\sim1$ than today.  This
point may apply to IC~10 X-1 and NGC 300 X-1.

\subsection{The van Kerkwijk model of photoionized winds in Cyg X-3 and the masses of black holes in Wolf-Rayet binaries}
\label{vanKerkwijk}

In recent years, attempts have been made to measure radial velocity
curves for extragalactic high mass X-ray binaries by following their
emission lines around the system orbit (Prestwich et al. 2007;
Silverman \& Filippenko 2008; Crowther et al. 2010).  While the
measurements may be correct, emission line tracers of orbital motions
in X-ray binaries are often unreliable.  In Cygnus X-3, it has been
shown that the radial velocities estimated from infrared emission
lines have their maximum blueshift at the X-ray and infrared
photometric minimum, and show their maximum redshift half an orbit
further along in phase (van Kerkwijk 1993).  A natural interpretation
of the relative phasing of the photometric and spectroscopic time
series is that the X-ray source photo-ionizes the entire stellar wind,
except for the component which is shielded by the donor star itself.
As a result, the radial velocity curve of the wind line does not trace
out the center-of-mass velocity of the star, but a particular
convolution of the center-of-mass velocity and the wind speed itself.
Under the assumption that the wind speed is larger than the orbital
velocity (which, in fact, follows from the system being detached),
then the amplitude of velocity variations is, essentially, just giving
a value close to the wind speed, and hence is a reliable probe of
neither the radial velocity of the donor star nor of the mass of the
compact object.

That the same phenomenon would take place for the cases of IC~10~X-1
(Prestwich et al. 2007; Silverman \& Filippenko 2008) and NGC 300~X-1
(Crowther et al. 2010) does not follow immediately from its taking
place for Cygnus X-3.  The X-ray luminosities of all three systems are
about the same, but the orbital period of Cygnus X-3 is a factor of
about 7 shorter, meaning that the system separations are likely to be
smaller as well, and the ionization parameter of the wind may be
smaller for the extragalactic systems than for Cyg X-3.  Nonetheless,
there are hints in both cases that this phenomenon is, in fact,
relevant to the two extragalactic systems.  In particular, in both
systems, the emission lines have a larger full-width half-maximum for
the phases where they have nearly zero central velocity than for the
phases near maximum blue shift or redshift (see Table 1 of Silverman
\& Filippenko 2008, and Figure 1 of Crowther et al. 2010).  A more
definitive determination of whether the lines trace out the stellar
wind velocity or the orbital velocity or some combination of the two
could be obtained by taking several spectra close enough in time to a
long X-ray observation that one could test the relative phasing of the
two time series.

\section{Summary}

We have presented evidence that the X-ray source
CXOU~J004732.0-251722.1 in the galaxy NGC~253 is likely to be a Wolf
Rayet X-ray binary -- the 4th member of the class discovered to date.
The evidence consists of the discovery of large amplitude X-ray
variability from the source which is consistent with being periodic on
a timescale of about 14-15 hours, and the source's location in the
center of a nuclear starburst galaxy.  We have also discussed how the
detections of individual Wolf-Rayet X-ray binaries may have
implications for the expected rate of gravtiational wave source
detections, and have illustrated both a path to using these sources
for such a purpose and several of the key challenges that will be
faced in applying that methodology.

\section{Acknowledgments}

TJM thanks Rob Fender for pointing out that the issues that apply to
the use of emission lines to measure a radial velocity curve for
Cygnus X-3 may apply to IC~10~X-1 and NGC~300~X-1 as well, and Fabien
Gris\'e for useful discussions about Wolf Rayet X-ray binaries.  TJM
also thanks the Astrophysics of the Canary Islands for hospitality
while this work was finished.  This work has made use of quick look
data provided by the ASM/RXTE team.  We thank an anonymous referee
whose comments have directed us to strengthen some of the analysis in
the paper and to improve the clarity and quality of several of the
figures.

\label{lastpage}

\end{document}